\documentclass[10pt]{article}
\usepackage{amsmath,amssymb,cite,epsfig,graphics,graphicx}

\numberwithin{equation}{section}

\begin{document}
\begin{flushright}
IISc-CHEP/6/04
\end{flushright}
\begin{center}
{\Large{\bf Quantum Topology Change and Large $N$ Gauge Theories}}
\bigskip 

Luiz C. de Albuquerque$^{\dag}$\footnote{lclaudio@fatecsp.br}, Paulo
Teotonio-Sobrinho$^\ddag$\footnote{teotonio@fma.if.usp.br} and Sachindeo
Vaidya$^\star$\footnote{vaidya@cts.iisc.ernet.in} \\ 
\bigskip
{\it $^\dag$Faculdade de Tecnologia de S\~ao Paulo - DEG - CEETEPS - 
UNESP, \\ Pra\c{c}a Fernando Prestes, 30, 01124-060 S\~ao Paulo, SP,
Brazil} \\
\bigskip
{\it $^\ddag$Universidade de S\~ao Paulo, Instituto de F\'{\i}sica -
  DFMA, \\
Caixa Postal 66318, 05315-970, S\~ao Paulo, SP, Brazil} \\
\bigskip
{\it $^\star$Centre for High Energy Physics, Indian Institute of Science, \\
560012, Bangalore, India}. 
\end{center}

\begin{abstract}
We study a model for dynamical localization of topology using ideas from
non-commutative geometry and topology in quantum mechanics. We consider a
collection $X$ of $N$ one-dimensional manifolds and the corresponding set
of boundary conditions (self-adjoint extensions) of the Dirac operator
$D$. The set of boundary conditions encodes the topology and 
is parameterized by unitary matrices $g_N$. A particular geometry is described 
by a spectral triple $x(g_N)=(A_X,{\cal H}_X, D(g_N))$. We define 
a partition function for the sum over all $g_N$. In this model 
topology fluctuates but the dimension is kept
fixed. We use the spectral principle to obtain an action for the
set of boundary conditions. Together with invariance principles the
procedure fixes the partition function for fluctuating topologies. In the
simplest case the model has one free-parameter $\beta $ and it is
equivalent to a one plaquette gauge theory.  We argue that topology becomes
localized at $\beta=\infty$ for any value of $N$.  Moreover, the system
undergoes a third-order phase transition at $\beta=1$  for large $N$. 
We give a topological interpretation of the phase transition by
looking how it affects the topology.
\end{abstract}

\section{Introduction}
A coherent picture embracing both quantum theory and gravity has been
a big challenge for over seventy years \cite{reviews}.  The last
decades have witnessed a conceptual change on the usual notions of
space-time and quantum mechanics.  It is generally agreed that at very
high energies the conventional ideas about the space-time breaks down,
so that the geometrical framework of General Relativity becomes
inadequate to describe the non-manifold micro-structure of space-time.
In string theory, for instance, $X_\mu$ are operators that happen to
be interpreted as coordinates of an embedding in a metric space
\cite{strings}.  Many theories also suggest a discrete picture of the 
space-time at very small distances. Thus, in loop quantum gravity the
operators of spatial area and volume have discrete spectra
\cite{RevLQG}. Besides, both string theory and loop quantum gravity
strongly indicate a non-commutative structure of the space-time at the
Planck scale \cite{RevLQG,NCST1}.

Although a complete, non-perturbative theory of quantum gravity is
still unknown it is fair to say that some theoretical progress has
been achieved.  This advance stimulated the development of a growing
quantum gravity phenomenology.  It is now argued that space-time
fluctuations at the scale of quantum gravity may be probed/tested at
energies accessible experimentally, now or in a near future.  Many
experimental proposals rely on departures from the classical Lorentz
symmetry due to space-time quantum fluctuations, with different
scenarios predicting similar modified dispersion relations 
\cite{Ame}. Possible tests include time-of-arrival difference between 
high-energy photons from gamma-ray bursts and observations of
high-energy cosmic rays above the GZK bound \cite{Ame}.
Other interesting experimental set-ups involve analysis of noise in
gravity-wave  \cite{Ame} and matter interferometers \cite{Per}. 
Besides, there is some input from current experimental data 
\cite{Bill,Lisi,Sud} which can place
constraints on the possible scales of space-time fluctuation
effects. Finally,  cosmological observables such as the cosmic microwave
background spectrum may also contain clues to the quantum structure of
spacetime \cite{CMB1,Lizzi}.

Any consistent theory of quantum gravity should, in the low energy
limit, give us the conventional geometric picture of space-time. In
such theory the space-time itself will be dynamically generated. Thus
its dimension, the signature of the metric, global topology, causal
structure, etc., will be computable (at least in principle)
observables of the theory, and not predetermined inputs. However,
almost all models assume a given dimensionality, signature and/or
topology from the beginning. This is the case in the simplicial approach
(quantum Regge calculus \cite{Regge2}, dynamical \lq\lq
triangulations" \cite{simplicial}, Lorentzian
triangulations\footnote{\footnotesize It is argued that the \lq\lq
effective" dimension of the space-time (associated with the ensemble
of random triangulations in the continuum limit) may be different from
the dimension of the underlying simplex. For instance, the Hausdorff
dimension ($d_H$) of 2D pure Euclidean gravity turns out to be four in
the dynamical triangulation approach \cite{Amb2}, in contrast with the
result $d_H=2$ in the Lorentzian triangulation \cite{DLT}. However,
this is seem somewhat as a \lq\lq pathology" of the Euclidean
formulation \cite{DLT}.}\cite{DLT}, etc.), and in the (perturbative)
string scenario where the target space has a fixed dimensionality to
be determined later by consistence conditions (e.g. anomaly
cancellations). In fact, string calculations assume a fixed background
classical space-time: it is hoped that a non-perturbative approach
would introduce manifest background independence into the theory, but
this remains a conjecture. Loop quantum gravity is already background
independent, however specific models are usually set from the start in
a 3+1 space-time.  There is some progress towards these opens
questions in other proposals. For instance, in the causal set theory
\cite{Sorkin} one starts from a minimum data input: a discrete set of
events endowed with a causal relation. A poset structure is also shown to
arise naturally in the so-called causal spin networks \cite{FS}.

Many theories suggest that the background of quantum gravity is
well described by some type of space-time foam \cite{Wheeler,Haw},
notably spin foam models and related ones. Thus, not only the
geometrical properties of the space-time, but its topology is also 
subject to quantum fluctuations at the quantum gravity
scale. There is some phenomenological proposals to uncover possible
macroscopic signals of quantum topology fluctuations, see for instance
\cite{Garay}.  It is widely believed that topology changes are pure
quantum phenomena, and a necessary ingredient for a consistent theory
of quantum gravity. For instance, there is no spin-statistics theorem
for geons (i.e. soliton-like excitations of a spatial manifold) unless
topology changing processes are allowed \cite{DS,geon2}. In this
paper we will study a toy model for topology fluctuations where it
will be possible to address dynamical questions in a simpler context.

Since the manifold structure of space-time has to appear at some
macroscopic limit, it is natural to expect that one needs a
generalization of ordinary geometry, such as noncommutative geometry
(NCG), to approach the Planck scale physics. The starting point of NCG
\cite{NCG} is the remarkable observation that one can describe a
Riemannian manifold $(M,g_{\mu \nu})$ in a purely algebraic way.
There is no loss of information if, instead of the data $(M,g_{\mu
\nu})$, one is given a triple $({\cal A},{\cal H}, D)$, where ${\cal
A}$ is the C*-algebra $C^0(M)$ of smooth functions on $M$, ${\cal H}$
is the Hilbert space of $L^2$-spinors on $M$, and $D$ is the
Dirac operator acting on ${\cal H}$. From the Gelfand-Naimark theorem
it is known that the topological space $M$ can be reconstructed from
the set $\widehat{\cal A}$ of irreducible representations of $C^0 (M)$.
Metric is also encoded, and the geodesic distance can be computed from
$D$.  In particular one can treat all Hausdorff topological
spaces in this way. Given a pair $(M,g_{\mu \nu})$, one can promptly
construct the corresponding triple $(C^0 (M),L^2(M),D)$. 
However, not all commutative spectral triples come from a pair 
$(M,g_{\mu \nu})$. Nevertheless one can
always associate a Hausdorff space $M=\widehat {\cal A}$ to a commutative
spectral triple, where $\widehat {\cal A}$ denotes the set of irreducible
representations of ${\cal A}$. The space $M$ may not be a
manifold and the spectral triple has to be regarded as a generalized geometry.

The framework of NCG suggests a possible approach to quantum gravity
based on the so-called spectral principle \cite{cc,Landi}: Once we
trade the original Riemannian geometry for its corresponding
commutative triple we need a replacement for the Einstein-Hilbert
action $S_{EH}$.  The spectral action of Chamseddine and Connes
\cite{cc} is one possible candidate. It is a very simple function of 
the eigenvalues of $D$ and contains $S_{EH}$ as a dominant term.

Spectral actions can be written for any triple, regardless of
whether it comes from a manifold $(M,g_{\mu \nu})$ or not. In the
spectral geometry approach it is thus conceivable to write the partition
function
\begin{equation}\label{1.1}
Z=\sum_{x\in {\cal X}}\,e^{-S[x]},
\end{equation}
where the \lq\lq sum'' is over the set ${\cal X}$ of all possible
commutative spectral triples and $S$ depends on the spectrum of $D$. 
It includes all Hausdorff spaces and therefore all manifolds of
all dimensions.

In a previous Letter \cite{PRL} we introduced a simple discrete model
to quantum gravity based on a particular truncation of the sum in
(\ref{1.1}). We discretized (\ref{1.1}) by sampling the set ${\cal X}$
with finite commutative spectral triples $x=({\cal A},{\cal H}, D )$
where the commutative \mbox{C*-algebra} ${\cal A}$ has a countable
spectrum $ {\widehat {\cal A}}$. In this approach to discretization there
is no need to introduce a lattice or simplicial decomposition of the
underlying space. The approximation of ${\cal A}$ by a finite
dimensional algebra works even if the spectral triple does not come 
from a manifold. Thus, it gives us a generalization of ordinary discretizations
\cite{simplicial,Regge2,discrete}. The model describes the geometry of spaces
with a countable number $n$ of points, and is related to the Gaussian
unitary ensemble of Hermitian matrices: for fixed $n$ the operator $D$
is a $n\times n$ self-adjoint matrix.  The average number of points in the
universe $\langle n \rangle$, the expectation value $\langle\delta\rangle$ of
the dimension, and the metric are macroscopic observables of the theory, obtained
after some suitable average (coarse-graining) over the ensemble. We
showed that the discrete model has two phases: a finite phase with a
finite value of $\langle n\rangle$ and $\langle\delta\rangle=0$, and
an infinite phase with a diverging $\langle n\rangle$ and a finite
$\langle\delta\rangle\ne0$. The critical point was computed as well as
the critical exponent of $\langle n\rangle$. Moreover, an upper bound
for the order parameter $\langle\delta\rangle$ was found,
$\langle\delta\rangle\leq 2$. The discrete model is  a pre-geometric
one, in the sense that the continuum picture with its 
geometrical content emerges through a phase  transition.

In the present paper we elaborate on another discrete model
where the dimension will be kept fixed while the topology
fluctuates. Again, we will consider only degrees of freedom 
associated to pure gravity, i.e. coupling with matter degrees of
freedom will not be included. Relying on the framework developed
in \cite{bbms}, we consider a collection $X$ of $N$ intervals of
length $L$. For this set of one dimensional manifolds, the
momentum operator $P$ plays the role of the Dirac operator. 
The sum in (\ref{1.1}) will be over triplets
$x=(A_X,{\cal H}_X, D=P)$ where $A_X$ is the algebra of continuous
functions on $X$ and ${\cal H}_X=L^2(X)$.  In order to fix the
spectral triple, however, we have to consider the self-adjoint
extensions of $P$, i.e. boundary conditions (b.c.). These are labeled by
unitary matrices $g$.  Thus, we are lead to compute a partition
function over all self-adjoint extensions of $P$. According to
Balachandran et al. \cite{bbms} the b.c. fixes the global topology of
the configuration space. The topology depends on the form of $g$, and
in general is different from the classical one.  In particular, it can
be a superposition of circles $S^1$ of different sizes. The definition
of the triplets and a short revision of the main arguments in
\cite{bbms} are the subject of Section {\bf 2}. In Section {\bf 3} we
use the spectral principle as a guide to obtain an effective action for
the $g$'s. This, together with symmetry requirements, fixes the
partition function.  Once we have the partition function for the
ensemble of all topologies  we are able to study the
dynamical localization of topology. This is done in Sections {\bf 4}
and {\bf 5}, where our main results are discussed. We identify the simplest
version of the model (with only one parameter, $\beta$) with the
Gross-Witten model which arises from the Wilson's lattice version of
YM$_2$. Namely, the partition function reduces to a generalization
of the Dyson's circular unitary ensemble.  We
numerically verify that the configuration space is a collection of
circles of size $L$, $S^1\cup S^1\cup\cdots \cup S^1$, for all finite
$N$ at $\beta=\infty$. Topology thus gets localized in this limit. 
In the large $N$ limit there is a phase transition at $\beta=\beta_c$.  
We also give a topological interpretation of the phase transition by
looking how it may affects the topology.

\section{Fluctuating Topology} 

The connections between topology and quantum mechanics have been
clearly exposed in \cite{bbms}. Here we rephrase the discussion using
the language of non-commutative geometry.

Let us consider a collection $X$ of $N$ one dimensional manifolds
(intervals) of length $L$. The corresponding spectral triple will be
taken as $x=(A_X,{\cal H}_X, D)$ where $A_X$ is the
algebra of continuous functions on $X$ and ${\cal H}_X=L^2(X)$.  The
analogue of the Dirac operator $D$ will be the momentum operator
$P$.

Let us consider a simple example where $X$ is a pair of disjoint
intervals $I_1, I_2$. The intervals will be parametrized by a
coordinate $x\in [0,L]$. The classical configuration space of a
particle living on $X$ is just the union $[0,L] \cup [0,L]$. An
element $\psi \in {\cal H}_X$ is a pair of functions $\psi _1(x),\psi
_2(x)$, \mbox{$\psi _i:I_i\rightarrow \mathbb{C}$} and the scalar
product is

\begin{equation}\label{2.1}
(\psi,\chi) = \int_0^L dx \sum_{i=1}^2 (\psi^*_i \chi_i)(x)
\end{equation} 
We write the wave-function conveniently as a column vector $\psi (x)=
(\psi_1 (x), \psi_2 (x))^t$ so that the operator $D=P_2$ takes the 
following matrix form
\begin{equation}
P_2=
\left(
\begin{array}{cc}
  -i\partial _x & 0\\
  0 & -i\partial _x 
\end{array}
\right)\label{2.2}
\end{equation}
We have not fixed completely the spectral triple. The operator $D$ is
fixed only up to boundary conditions (b.c.) or self-adjoint
extensions. Let the eigenfunctions of the operator $P_2$
be of the form
\begin{equation} \label{2.3}
\psi(x) = \left(\begin{array}{c}
              A \\
              B
          \end{array}\right) e^{i p x},
\end{equation} 
where $A, B$ and $p$ are obtained by solving the equation
\begin{equation} \label{2.4}
\psi(L) = g \psi(0),
\end{equation} 
with $g \in U(2)$ parameterizes the b.c. or self-adjoint extensions.

One may ask what geometrical properties of $X$ are determined by such
b.c.. The point of view taken by Balachandran et all. in \cite{bbms}
is that a b.c. fixes the global topology.  Depending on the form of
$g$, the topology perceived by the quantum particle is quite different
from the classical one.  Let us look at a couple of examples to
clarify this:

\begin{eqnarray} \label{2.5}
(a) \quad g_a &=& \left(\begin{array}{cc}
                   0 & e^{i \theta_{12}} \\
                   e^{i \theta_{21}} & 0
                 \end{array} \right), \\
(b) \quad g_b &=& \left(\begin{array}{cc}
                   e^{i \theta_{11}} & 0 \\
                   0 & e^{i \theta_{22}}
                 \end{array} \right)\label{2.6}
\end{eqnarray} 
The probability of finding the particle on the first interval is $\int
\psi^*_1 \psi_1 dx$, and similarly for the second interval. In the
case (a), the density functions $\psi_i^* \chi_i$ satisfy the
conditions
\begin{eqnarray} \label{2.7}
(\psi^*_1 \chi_1)(L) &=& (\psi^*_2 \chi_2) (0), \\
(\psi^*_2 \chi_2)(L) &=& (\psi^*_1 \chi_1)(0)\label{2.8}
\end{eqnarray} 
In other words, the probability densities are the same at the points
joined by the thin line (Fig.1), and thus the configuration space of
the particle is a circle made by joining the two intervals. The
eigenfunctions (\ref{2.3}) are of the form ($A_{\pm}=\pm\,{\rm
exp}\{i(\theta_{12}-\theta_{21})/2\}$)
%-----------------Figura 1-------------------------------------
\begin{figure}[hbt]
\begin{center}
\begin{picture}(0,0)(0,0)    
\put(15,-10){$0$}
\put(140,-10){$L$}
\put(60,-10){$L$}
\put(90,-10){$0$}
\end{picture}
\includegraphics[scale=0.5]{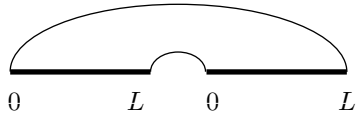}
\end{center}
\caption{
\footnotesize
The figure shows a boundary condition corresponding to a 
circle of size $2L$.}
\label{fig:1}
\end{figure}
%------------------------------------------------------------
\begin{eqnarray} \label{2.9}
& & \psi_n^{(+)}(x) = \left(\begin{array}{c}
                   A_{+} \\
                   1
          \end{array} \right) e^{i (n + \frac{\theta_{12} +
                   \theta_{21}}{4 \pi})\frac{2 \pi x}{L}},\\
& & \psi_n^{(-)}(x) = \left(\begin{array}{c}
                   A_{-} \\
                   1
          \end{array} \right) e^{i (n + \frac{\theta_{12} +
                   \theta_{21}}{4 \pi}+\frac{1}{2})\frac{2 \pi x}{L}},
\label{2.10}
\end{eqnarray}
and the spectrum is the set $\{\frac{2\pi}{L}\left(n +
\frac{\theta_{12} + \theta_{21}}{4\pi}\right)\}\cup\{\frac{2\pi}{L}
\left(n + \frac{\theta_{12} +
\theta_{21}}{4\pi}+\frac{1}{2}\right)\}$, $n \in {\bf Z}$.

In the case (b), the probability densities instead satisfy
\begin{eqnarray}\label{2.11} 
(\psi^*_1 \chi_1)(L) &=& (\psi^*_1 \chi_1)(0), \\
(\psi^*_2 \chi_2)(L) &=& (\psi^*_2 \chi_2)(0)\label{2.12}
\end{eqnarray} 
Now, as is obvious from Fig.2, the underlying configuration space is
the union of two circles.
%-----------------Figura 2-------------------------------------
\begin{figure}[hbt]
\begin{center}
\begin{picture}(0,0)(0,0)    
\put(20,-10){$0$}
\put(85,-10){$L$}
\put(122,-10){$0$}
\put(192,-10){$L$}
\end{picture}
\includegraphics[scale=0.7]{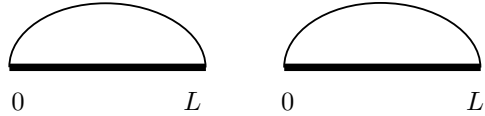}
\end{center}
\caption{
\footnotesize
The figure shows a boundary condition corresponding to a pair of
circle of sizes $L$.}
\label{fig:2}
\end{figure}
%------------------------------------------------------------

The eigenfunctions are of the form 
\begin{eqnarray} \label{2.13}
\psi_n^{(u)} (x) &=& \left(\begin{array}{c}
                         1 \\
                         0
                       \end{array} \right) e^{i (n +
                       \frac{\theta_{11}}{2\pi})\frac{2\pi x}{L}}, \\
\psi_n^{(d)} (x) &=& \left(\begin{array}{c}
                         0 \\
                         1
                       \end{array} \right) e^{i (n +
                         \frac{\theta_{22}}{2\pi})\frac{2\pi x}{L}},
\label{2.14}
\end{eqnarray} 
so that ${\rm Spec}(P)= \{\frac{2\pi}{L}\left(n+
\frac{\theta_{11}}{2\pi}\right)\} \cup \{\frac{2\pi}{L}\left( n +
\frac{\theta_{22}}{2\pi}\right) \}$.

For most other choices of $g$, topology is not localized as in these
two examples but it is rather a superposition of both, and there is no
classical interpretation to it. This happens also for other unitary
matrices corresponding to non-trivial b.c..

Notice that the spectrum depends only on the eigenvalues of the matrix
$g$. This is not merely a coincidence for the examples we looked at
here. It is easy to see that two matrices $g_1$ and $g_2$ that are
related by $g_2 = u g_1 u^\dag$ give rise to the same spectrum.

The generalization to arbitrary number $N$ of intervals is
straightforward. Our interest is in the operator
\begin{equation}\label{2.18} 
P_N = \left(\begin{array}{cccc}
        -i \partial_x & 0 & \cdots & 0 \\
           0 & -i\partial_x & \ddots & \vdots \\
           \vdots & \ddots & \ddots& 0 \\
           0 & \ldots & 0 & -i\partial_x
            \end{array} \right)
\end{equation} 
The self-adjoint extensions are labeled by a unitary $N \times N$
matrix $g$. The topology of the configuration space, as seen by the
quantum particle, is dictated by $g$. Different choices of
$g$ can give rise to, for example, a single classical circle or $k$ ($k<
N)$ disjoint circles. The spectrum of $P_N$ may be written as
$\{\frac{2\pi}{L}(n+\frac{\alpha_1}{2\pi}) \} \cup
\{\frac{2\pi}{L}(n+\frac{\alpha_2}{2 \pi}) \} \cup \cdots
\{\frac{2\pi}{L}(n+\frac{\alpha_N}{2 \pi}) \}$ where $(e^{i\alpha_1},
e^{i\alpha_2}, \ldots, e^{i\alpha_N})$ are the eigenvalues of the
matrix $g$.

The set of all possible matrices $g$ describes the set of 
topologies that the quantum particle sees. As remarked before,
only a small subset have a classical interpretation, since
classical topology corresponds to isolated points on the group
manifold.
Is there any natural sense in which one can
associate a probability to a particular topology i.e. for the matrix
$g$ from this ensemble? In other words, is it possible to write down
some kind of partition function for the $g$'s? A first step towards
this direction was done in \cite{bbms}, where a dynamics for the b.c.
(quantized boundary conditions) was proposed in the connection
picture. Here we follow another route, along the ideas introduced in
\cite{PRL}. Thus, we would like to compute the partition function
\begin{equation}
Z_N=\int [dg]\, e^{-S[x(g)]},\label{2.19}
\end{equation}
where $x(g)=(A_X, H_X, P_N(g))$. Next we
look at the probability distribution for the b.c., that is for the
unitary matrices $g$.  Then we will be able to ask questions on a
possible dynamical localization of topology.

\section{Spectral Action}

In order to compute (\ref{2.19}) we need to specify a dynamics,
i.e. determine an action $S[x]$ for the triple $x$.  Our guide will be
the spectral principle introduced in \cite{cc}.  Their proposal for
the action is, loosely speaking, just the trace of some positive
function of the square of the Dirac operator. In our case, this would
imply using the action
\begin{equation}\label{3.1} 
S_N = {\rm Tr}\, \chi\left(\frac{P_N^2}{\Lambda^2}\right)~,
\end{equation} 
where $\Lambda\equiv 1/L_\Lambda$ is a momentum cut-off.  The trace
class function $\chi(x)$ is typically chosen to be 1 for $x<1$ and
smoothly going to zero for $x>1$. It turns out that $S_N$ is
proportional to the number of eigenvalues with absolute value less
than $\Lambda$, $S_N\sim N n(\Lambda)$. Let ${\cal P}=(L/2\pi)\,P$ and
$\epsilon=(2\pi L_\Lambda/L)^2$. Most of the contribution to the sum
in (\ref{3.1}) comes from modes with $n+\alpha/2\pi$ less than or of
order $1/\sqrt{\epsilon}\sim L/L_\Lambda$, whereas higher modes make
almost no contribution.  Thus, one naively gets

\begin{equation} \label{3.3}
S_N (\alpha; \epsilon)\sim N\sum_{n} \chi\left(\epsilon
(n+\frac{\alpha}{2\pi})\right) \sim N\sum_{|n|}^{ 1/\sqrt{\epsilon}}1
\sim N\frac{L}{L_\Lambda}~. 
\end{equation}
As expected, $S_N\to0$ for $\epsilon\to\infty$ at fixed $N$. This is
natural since for $L_\Lambda\to\infty$ we are effectively cutting-off
all modes.

A regularized action that maintains the invariance $\alpha_k
\rightarrow \alpha_k + 2 \pi$ comes from adopting
$\chi(P_N^2/\Lambda^2) = e^{-\epsilon {\cal P}_N^2}$, giving

\begin{equation} \label{3.2}
S_N (\alpha_i; \epsilon)=\sum_{k=1}^N \sum_{n_k=-\infty}^\infty 
e^{-\epsilon(n_k +
    \frac{\alpha_k}{2\pi})^2}.
\end{equation}

We are concerned with the heat-kernel expansion of $S_N$ in
(\ref{3.2}) for $\epsilon\to0$. This follows at once from the modular
transformation,

\begin{equation} \label{3.4}
\sum_{n=-\infty}^\infty e^{-t(n+z)^2} = \sqrt{\frac{\pi}{t}} \left(1 +
2\sum_{n=1}^\infty e^{-\pi^2 n^2/t} \cos(2 \pi n z) \right),
\end{equation} 
so that the regularized action reads 
\begin{equation} \label{3.5}
S_N (\alpha; \epsilon)= \sqrt{\frac{\pi}{\epsilon}}\left[
N+2\sum_{k=1}^N \sum_{n_k=1}^\infty e^{-\pi^2
    n_k^2/\epsilon} \cos (n_k \alpha_k) \right], 
\end{equation} 
and one obtains an expansion for the effective action in the form

\begin{equation}\label{3.7} 
S_N (\alpha_i; \epsilon) = a_0 (\epsilon) + 
a_1 (\epsilon) \left(\sum_{k=1}^N\cos
\alpha_k\right) + a_2 (\epsilon) \left(\sum_{k=1}^N \cos (2\alpha_k)\right)
+ \ldots
\end{equation} 
Besides, using ${\rm Tr}(g+g^\dag)=2 \sum_k \cos \alpha_k, {\rm
  Tr}(g+g^\dag)^2 = 2N + 2\sum_k \cos 2\alpha_k,$ etc. we can re-write
  the spectral action in terms of the matrix $g$ itself:

\begin{equation} 
S_N (g; \epsilon) = b_0 (\epsilon) + b_1 (\epsilon) {\rm Tr} (g +
g^\dag) + b_2 (\epsilon) (g+g^\dag)^2 +\ldots
\label{3.8}
\end{equation}  

To leading order, apart from an overall additive (and hence
irrelevant) constant $b_0(\epsilon)$, this is nothing but Wilson's
action for a 2-d Yang-Mills gauge theory on a single
plaquette \cite{wilson,gw}. Including
higher order terms gives us models of the type considered in
\cite{ps}.

As we noted earlier, the spectrum of the operator $P_N$ is unchanged
when the matrix $g$ is conjugated by a unitary matrix $u$. We will
require our action and the corresponding partition
function to have the same invariance. The partition function
(\ref{2.19}) is thus of the form
\begin{equation}\label{3.9}
Z_N (b_\ell) = \int [dg] e^{-S_N [g,b_\ell]} 
\end{equation} 
where $[dg]$ is the $U(N)$-invariant Haar measure on the group $U(N)$.
In terms of the eigenvalues $e^{i \alpha_j}$ of the matrix $g$, the
partition function becomes \cite{Mehta}
\begin{equation}\label{3.10} 
Z_N (b_\ell) = \int_0^{2\pi} [d \alpha_j]
\Delta(\{\alpha_i\}) \bar{\Delta}(\{\alpha_i\}) e^{- S_N
  (\alpha_k,b_\ell)},
\end{equation} 
where $\Delta(\{\alpha_i\})$ is the Vandermonde determinant
\begin{equation} \label{3.11}
\Delta(\{\alpha_i\}) = \prod_{i<j} (e^{i \alpha_i} - e^{i \alpha_j}),
\end{equation} 
and the normalization is $Z_N(b_l=0)=1$:
\begin{equation}\label{3.12} 
[d \alpha_j] \equiv \frac{1}{N!(2\pi)^N}\prod_{j=1}^N  d \alpha_j,
\end{equation} 

\section{Localization of Topology}
Let us restrict our attention to the simplest non-trivial truncation
of (\ref{3.8}), which we write as
\begin{equation} 
S_N (g,\beta) = \frac{N\beta}{2}{\rm Tr} (g + g^\dag),
\label{4.1}
\end{equation} 
where the factor $N/2$ is for later convenience. 
The partition function reads
\begin{equation}\label{4.2} 
Z_N (\beta) = \int [d \alpha_k]
e^{-{\cal H}_N(\alpha_i,\beta)},
\end{equation} 
where

\begin{equation}\label{4.3} 
{\cal H}_N={N\beta}\sum_{k=1}^N \cos\alpha_k-2\sum_{i<j}
\ln|e^{i \alpha_i} - e^{i \alpha_j}|.
\end{equation}

The action (\ref{4.1}) has been extensively studied in the literature
in connection with YM$_2$, and has interesting properties in the large
$N$ limit \cite{gw}. Besides, $Z_N(0)$ is the matrix integral
of the Dyson's circular unitary ensemble.  For finite $N$, using the
identity \cite{Forrester}
\begin{equation}\label{4.4} 
\frac{1}{N!}\int_0^{2\pi} \prod_{k=1}^N d\alpha_k\,\prod_{\ell=1}^N
g(\alpha_\ell)\, 
\prod_{i<j} |e^{i \alpha_i} - e^{i \alpha_j}|^2=
\det \left[\int_0^{2\pi}d\alpha g(\alpha)
  e^{i\alpha(i-j)}\right]_{i,j=1,\cdots N}, 
\end{equation}
with $g(\alpha)=e^{-N\beta\cos\alpha}$, it is easy to show that

\begin{equation}\label{4.5} 
Z_N (\beta) =
\det\left[I_{|i-j|}\left({N\beta}\right)\right]_{i,j=1,\cdots
  N}, 
\end{equation}
where $I_\nu(x)$ is the  modified Bessel function of the first kind.

Let us sum up the argument developed up to now.  We are considering a
collection of $N$ disjoint compact 1D manifolds. A point $x$ in this
collection is a  union of circles and intervals with some b.c., 
corresponding to a particular self-adjoint extension of the momentum operator.
Accordingly, to each element $x$ we assign a given topology,
parameterized by a unitary matrix $g\in U(N)$ (that is, a boundary
condition).  Since in our model $Z_N$ is the partition function for
the set of all self-adjoint extensions of the momentum operator,
i.e. all points $x$, it is sensible to interpret

\begin{equation}\label{4.6} 
P_N (x) = \frac{e^{-S[x]}}{Z_N},
\end{equation}
as the probability of having a configuration space with the topology
of $x$.  It is clear from Section {\bf 2} that the topology will be in
general \lq\lq fuzzy", i.e., it may not admit a classical
interpretation.  Now with model (\ref{4.2}) we want to ask
questions such as : is it possible that the topology gets localized
around a classical configuration for some value of $\beta$? 
In other words, is it possible that
$P_N(x,\beta)$ gets localized around some particular manifold $x$?

We stress that the topology of $x$, or the boundary condition (\ref{2.4})
is determined by $g$, but the probability measure in (\ref{4.2}),

\begin{equation}\label{4.6.b} 
P_N (g,\beta) = \frac{1}{Z_N(\beta)}
\frac{e^{-{\cal H}_N(\alpha_i,\beta)}}{N!(2\pi)^N},
\end{equation}
depends only on the eigenvalues $e^{i\alpha_k}$ of $g$.  It does not
picks out a topology but rather an orbit of $g$ under conjugation.  In
other words, there is not a one-to-one correspondence between the set
of all topologies and the eigenvalues of $g$, as mentioned in Section
{\bf 2}. The only exception is the identity matrix ${\rm
Spec}(g)=\{1,1,\cdots 1\}$, which corresponds to the disjoint union $S^1\cup
S^1\cup S^1\cdots \cup S^1$ of circles of size $L$. Thus only in
this case one may speak of a dynamical localization of topology.

Notice that $Z_N$ may be interpreted as the partition function (at
fixed temperature $T$) of a 1D plasma of equal charged point-particles
constrained to move in a thin circular wire of radius one immersed in
a 2D world (plane). The second-term in the \lq\lq Hamiltonian" ${\cal H}_N$
is the 2-body repulsive Coulomb potential, whereas the first term
represents a periodic potential with strength given by $N\beta$. It
is well-known that at $\beta=0$ the system displays only a single
phase over all the temperatures scale, characterized by a long-range
order of crystalline type \cite{Dyson}. The spectral density
$\sigma_N(\alpha)=\left\langle\sum_k
\delta(\alpha-\alpha_k)\right\rangle_N/N$ is uniform around the unity
circle, $\sigma_N(\alpha,\beta\to 0)=1/2\pi$, and the topology is
\lq\lq fuzzy''.

However, it is conceivable that such situation does not hold at finite
$\beta$. Thus, at some value $\beta_c$ the strength of the periodic
potential may be enough to disorder the crystal structure, leading to
a melting of the Dyson crystal into a new phase.  This conjecture is
supported by a numerical analysis of some \lq\lq observables''. In
particular, by means of (\ref{4.5}) it is possible to shown
numerically that 

\begin{equation}\label{4.7}
\langle \cos\alpha_\ell\rangle_N(\beta)
=-\frac{1}{N^2}\frac{\partial}{\partial\beta}\ln Z_N (\beta)=
\left\{
          \begin{array}{ll}
          0 & \mbox{if $\beta\to 0$}, \\
          1 & \mbox{if $\beta\to \infty$}.
          \end{array}
        \right.
\end{equation}
For $\beta\to0$ the eigenvalues become uniformly distributed around
the unity circle, the topology is fuzzy, and $\langle
\cos\alpha_\ell\rangle_N\to0$ as expected. On the other hand, for 
$\beta\to\infty$ the periodic potential overcome the level repulsion
and the eigenvalues tend to concentrate around the origin, i.e.
matrices $g\sim \mathbb{I}$  are favored. 
This is a clear signature of a 
dynamical localization of topology at $\beta=\infty$ for any value of $N$.

\section{Topology and the Third Order Phase Transition}

Gross and Witten have shown that the one-plaquette model
described by (\ref{4.1}) and (\ref{4.2}) undergoes a third order phase
transition at $\beta = 1$ in the large $N$ limit \cite{gw}. In this
section we would like to discuss whether this phase transition has any
consequences to classical topology in our model of fluctuation
topologies. One has to keep in mind that topology is
described by the matrix $g_N$ of  boundary conditions. However, the
dynamics depends only on the eigenvalues of $g_N$. In other words,
a single set of eigenvalues determines a submanifold of boundary
conditions. Since the
model is not very sensitive to the topology, we do
not expect strong topological changes as we tune $\beta $ across the
critical point. The only point where topology is sharply affected is for
$\beta \rightarrow 0$ as explained in the last section. Nevertheless,
it is possible to give a topological interpretation for the phase
transition.

Let us summarize the results we need from \cite{gw}. For large
values of $\beta$ (weak coupling), the density of eigenvalues
$\sigma(\alpha,\beta)$ is strongly peaked near $\alpha=0$, whereas
the density is almost uniform over the unit circle for $\beta \simeq
0$ (strong coupling).  More precisely, each phase is
characterized by an appropriate spectral density (we change the domain
of $\alpha$ from $[0,2\pi]$ to $[-\pi,\pi]$)

\begin{eqnarray} \label{4.8} 
& &\sigma(\alpha,\beta)=\frac{1}{2\pi}\left(1+{\beta}
\cos\alpha\right),\qquad -\pi\leq\alpha\leq\pi;\\ 
& &\sigma(\alpha,\beta)=\frac{\beta}{\pi} \cos\frac{\alpha}{2}\,
\sqrt{\beta^{-1}-\sin^2\frac{\alpha}{2}},\qquad
-\alpha_c\leq\alpha\leq\alpha_c,\qquad
\sin^2\frac{\alpha_c}{2}=\beta^{-1}\,,\label{4.9} 
\end{eqnarray}   
valid for $\beta\leq1$ and $\beta\geq1$, respectively. The signature
of the phase transition at $\beta_c=1$ is clear: for $\beta>>1$ the
spectral density has support at a small region around $\alpha\simeq0$,
and the probability of finding an eigenvalue outside of this region is
zero. As we decrease $\beta$ the support of $\sigma_N(\alpha,\beta)$
becomes a larger arc of the unity circle around $\alpha=0$. For
$\beta>\beta_c$ there always be a gap (forbidden region in the
eigenvalues space) on the unity circle around $\alpha=\pi$. The gap is
closed at $\beta=\beta_c$.

Consider a matrix $g$ of the form

\begin{equation} \label{4.10}
g_N=\left(\begin{array}{cc}
        B(k) & 0 \\
           0 & S(N-k)
          \end{array} \right),
\end{equation} 
where $S$ is an arbitrary unitary and $B(k)$ is the $k\times k$ matrix
\begin{equation}\label{4.11}
B(k) = \left(\begin{array}{cccc}
        0 & w_1 & \cdots & 0 \\
           0 & 0 & w_2 & \vdots \\
           \vdots & \ddots & \ddots& w_{k-1} \\
           w_k & \ldots & 0 & 0
            \end{array} \right)
\end{equation}
with $w_i=e^{i\alpha_i}$. 
Thus, this b.c. has a big circle of size $k\,L$ as a classical
manifold.  The big circle is made of intervals number $1, 2, 3,\cdots,
k$. The remaining $(N-k)$ intervals are connected by an arbitrary
boundary condition given by $S$ and may not admit a classical 
interpretation. Furthermore, any other matrix that is related to
(\ref{4.11}) by a permutation also has a classical big circle of size 
$k\,L$.  
 
Let us call $C_k\subset U(N)$ the subset of boundary conditions of
type (\ref{4.10}) and its permutations.

The $k$ eigenvalues of $B(k)$ is a subset of ${\rm Spec}(g)$. 
To find them we write
\begin{equation}\label{A.2}
[B(k)]_{ij}=\omega_i\,\delta^j_{(i+1){\rm mod}(k)}.
\end{equation}
It follows that $B(k)^k=e^{i\gamma}\,\mathbb{I}$, where
$e^{i\gamma}=\omega_1\omega_2\cdots\omega_k$.  Therefore the
eigenvalues are $\lambda_m=\exp\left(i\frac{2\pi m+\gamma}{k}\right)$,
$m\in\{0,1,\cdots k-1\}$.  The eigenvector corresponding to
$\lambda_m$ is $v_m = (1,\frac{\lambda_m}{w_1}, \frac{\lambda_m^2}{w_1
w_2},\ldots, \frac{\lambda_m^n}{w_1 w_2 \ldots w_n} , \ldots,
\frac{w_k}{\lambda_m})$.
Notice that the eigenvalues of $B(k)$ are
equally spaced. They occur at the vertices of a regular polygon
inscribed in the unity circle.

The existence of a gap for  $\beta >\beta_c$ means that the probability
density has a support on a subset $H(\beta)\subset U(N)$ of all boundary
conditions. 
%-----------------Figura 3-------------------------------------
\begin{figure}[hbt]
\begin{center}
\begin{picture}(0,0)(0,0)    
\put(80,85){$\alpha_c$}
\end{picture}
\includegraphics[scale=0.5]{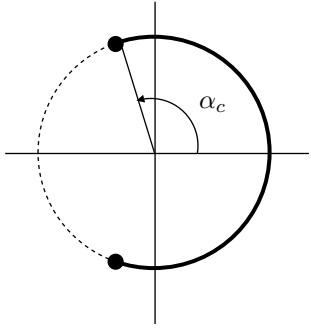}
\end{center}
\caption{
\footnotesize
The support of the spectral density is represented on the circle as a
dark line interval determined by the angle $\alpha_c$.}
\label{fig:3}
\end{figure}
%------------------------------------------------------------

Let us consider  Fig.3. The dark region is the support of 
$\sigma_N(\alpha,\beta ^{-1})$ and $\alpha_c=\alpha_c(\beta^{-1})$ 
is an increasing function of $\beta ^{-1}$.  For $\beta ^{-1}\to 0$ 
we have $\alpha_c\approx0$. In other words, the largest classical 
big circle has size $k=L$.  Let us see what we need to have a 
classical big circle of size $k=2\,L$ inside the allowed region.  
The eigenvalues of $B(2)$ are $\exp\left(i\frac{\gamma}{2}\right)$ and
$\exp\left(i\frac{\gamma}{2}+i\pi\right)$.  Suppose we have
$\alpha_c=\frac{\pi}{2}+\epsilon$. Many classical circles of size 
$2\,L$ given by $B(2)$ will be suppressed. However the b.c. such that
$\frac{\pi}{2}-\epsilon\,<\,\gamma\,<\,\frac{\pi}{2}+\epsilon$ will be
allowed. For smaller values of $\alpha_c$, $k=1$ is the biggest 
classical circle possible. A similar argument shows that classical big circles 
with size $k=m\,L$ will only show up for
\begin{equation} \label{4.13} 
\alpha_c\,\geq\,\frac{m-1}{m}\pi\;.  
\end{equation}  

In other words, for a fixed $\beta$, if we look at the intersection 
of $H$ and $C_k$ we see that $H$ excludes all sets $C_k$ for $k$ 
larger than some value $k_{\rm max}$. 
In Fig.4 we plot $k_{\rm max}$ as a function of $\beta ^{-1}$.  The
graph looks like a staircase function.  However as we approach
$\beta_c$, the variable $k_{\rm max}$ tends to $\infty$ (in the
thermodynamic limit $N\to\infty$) and at the same time the sizes of the
plateaus go to zero.
\bigskip
  
%-----------------Figura 4-------------------------------------
\begin{figure}[hbt]
\begin{center}
\begin{picture}(0,0)(0,0)    
\put(155,0){$\beta^{-1}_c$}
\put(190,10){$\beta^{-1}$}
\put(10,157){$k_{\rm max}$}
\end{picture}
\includegraphics[scale=0.5]{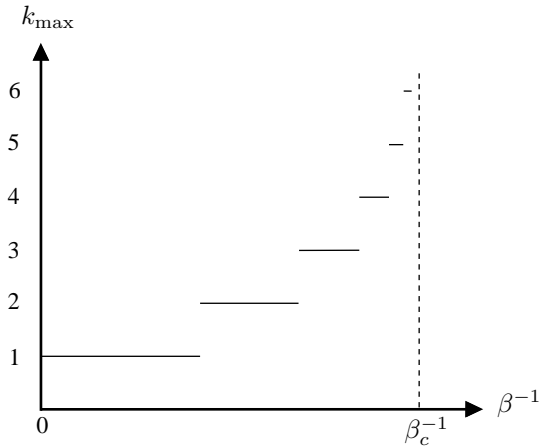}
\end{center}
\caption{
\footnotesize
Plot of $k_{\rm max}$ as a function of $\beta^{-1}$.}
\label{fig:4}
\end{figure}
%------------------------------------------------------------

We can also look at another quantity:
$\ell(\beta)=\frac{k_{\rm max}}{N}$.  In the limit $N\to\infty$,
\begin{equation} \label{4.14}
\ell(\beta)=
\left\{
          \begin{array}{ll}
          0 & \mbox{if $\beta\,\geq\,\beta_c$}, \\
          1 & \mbox{if $\beta\,<\,\beta_c$}.
          \end{array}
        \right.
\end{equation}
Thus we have the following qualitative picture of the phase transition
at $\beta_c$: above $\beta_c$ we only find classical circles of finite size.
There are other b.c. but all classical circles are at most $k_{\rm max}\,L$ in
size.  Below $\beta_c$ there are classical circles of arbitrary sizes. 
The order parameter is just $\ell(\beta)$.

\section{Conclusions}

The above model and the one in \cite{PRL} are not intended to be
realistic. In particular, time does not appear at all in our present
model. One can imagine that an equilibrium configuration has been
attained, in which the fluctuations in the dimension, metric and
signature of the space-time had already been partially localized.
Our model would thus be an effective
theory describing the topology fluctuations, controlled by the
parameter $\beta$. This is conceptually similar to the approach
followed in \cite{geon2}. Rather our intention here is to furnish a
hint on how NCG, in the formulation embodied in (1.1), may be used to
tackle some difficult, open questions in quantum gravity.  Thus, in
this paper we asked if dynamics can fix topology somehow. We
believe that these simple models capture some main features of more
elaborate ones, and hope that they could furnish insights into
it. This is in accordance with current views on universality in
quantum gravity \cite{simplicial,Fotini2}.  The idea of universality
is enforced here by the connection with random matrix theory. Thus, it
can be shown that the upper bound for the dimension observable 
$\langle\delta\rangle$ found in
\cite{PRL} does not change if instead of complex self-adjoint matrices
$D$ one considers real self-adjoint matrices, corresponding to the
Gaussian orthogonal ensemble \cite{unp}.  Besides, the key role played
by the eigenvalues of the Dirac operator in GR and in the spectral
action approach was emphasized in \cite{Landi}: they are
diffeomorphism-invariant functions of the metric and can be taken as
the dynamical variables of GR. In our model they are also the natural
dynamical variables due to the connection with random matrix theory.

Inspired by ideas from topology in quantum mechanics and relying
on the framework of the non-commutative geometry we have set-up
a simple model to study fluctuations in topology for a collection of
$N$ one-dimensional manifolds. In its simplest version the model has 
one free-parameter, $\beta$, and its partition function reduces to
the partition function of the one plaquette $U(N)$ gauge theory.
Although our simple dynamics is not particularly sensitive to the
topology of the underlying configuration space 
(since it depends only on the eigenvalues of the unitary matrices 
$g$ parameterizing the  boundary conditions), we have argued that 
topology gets localized at $\beta\to\infty$ for any value of $N$. 
For large $N$ the model
has a third-order phase transition at $\beta_c=1$. Topology is not, 
in general, localized for $\beta>\beta_c$ and large $N$, however
some topologies are excluded due to the finite support of the spectral
density in this range of $\beta$. Thus it seems possible that, in 
more realistic models, topology can be indeed fixed by 
the dynamics. 

The model discussed in the present work 
points to the remark that an eventual theory of quantum gravity
at the Planck scale may possible contain more degrees of freedom
than what one would naively expect based on the macroscopic
space-time physics \cite{bbms}. There are many possible ways to extend this 
and the related work \cite{PRL}. Notice that the present work 
and \cite{PRL} are somehow complementary: whereas in the latter 
we have studied fluctuations in the dimension,
here we have focused on topology fluctuations keeping the
manifold dimension fixed. Thus, it would be interesting to 
workout a model including both types of fluctuations,
with degrees of freedom associated with topology and
geometrical dynamics. Another possibility is to include
couplings with matter degrees of freedom. We hope that 
the toy models discussed here in connection with the
partition function (\ref{1.1}) set the stage towards 
its evaluation in a more realistic scenario where a phenomenological
approach can be eventually pursued.

\vspace{1cm}
{\bf Acknowledgments:}
We thank  A. P. Balachandran for suggestions and valuable comments.
L.C.A. and P.T.S. are grateful to all participants of the 
\lq\lq Workshop on Fuzzy Physics'', held at CINVESTAV in 
June 2004, for helpful
discussions which clarified some points of the work.
L.C.A. would like to thank the Mathematical Physics Department of 
USP at S\~ao Paulo for their kind hospitality. This work was 
partially supported by CNPq, grant 307843/2003-3 (L.C.A.).

\end{document}